# Effect of annealing on magnetic and magnetotransport properties of $Ga_{1-x}Mn_xAs$ epilayers

I.Kuryliszyn-Kudelska[a,*], T.Wojtowicz[a,b], X.Liu[b], J.K.Furdyna[b], W.Dobrowolski[a], J.Z.Domagala[a], E.Łusakowska[a], M.Goiran[c], E.Haanappel[c], O. Portugall[c]

[a]*Institute of Physics, Polish Academy of Sciences, Al.Lotników 32/46, Warsaw, PL-02-668, Poland*

[b]*Department of Physics, University of Notre Dame, Notre Dame, IN 46556, USA*

[c]*Laboratoire National des Champs Magnetiques Pulses Toulouse, 143 Avenue de Rangueil 31432 Toulouse Cedex 04, France*

**Abstract**

High-field magnetic measurements performed with the use of magnetooptical Kerr effect (MOKE) in the polar configuration as well as high-field and low-field magnetotransport studies were carried out on $Ga_{1-x}Mn_xAs$ epilayers grown by low temperature molecular beam epitaxy, and subsequently annealed under various conditions. The structural investigations by means of high resolution XRD were also performed. We observe significant changes in magnetoresistivity curves, magnetization and strain introduced by the annealing.

Ferromagnetic $Ga_{1-x}Mn_xAs$ continues to be the subject of intense interest due to its interesting physical properties as well as possible spin-electronics applications (see e.g. [1] and references therein). Our previous papers [2,3] showed that LT annealing rearranges Mn sites in $Ga_{1-x}Mn_xAs$ lattice. Annealing at the optimal conditions leads to the increase in $T_C$, saturation magnetization and free hole concentration due to the removing of Mn atoms from interstitial ($Mn_I$) positions [2,3]. Theoretical calculations [4] showed that $Mn_I$ does not contribute to the ferromagnetic coupling mediated by free holes and that $Mn_I$-$Mn_{Ga}$ pairs are coupled antiferromagnetically through the superexchange interaction. The aim of the present studies was to explore the structural, magnetic and magnetotransport properties for the as-grown layers and layers annealed at various conditions. The 100nm-400nm thick $Ga_{1-x}Mn_xAs$ layers were grown at the temperatures $T_S$~212°C-275°C with Mn concentration $x$ ranging from 0.014 to 0.083. The measurements of magnetoresistance were performed with the magnetic field applied perpendicular to the plane of the film. Static fields up to 13T and pulsed fields up to 55T were used. All samples indicate unsaturated negative magnetoresistance (*MR*) up to the highest value of investigated field. We found that annealing of the $Ga_{1-x}Mn_xAs$ epilayers with high Mn concentration leads to very significant changes in magnetoresistivity. Fig.1 shows typical behavior of *MR* for the sample with high Mn content $x$=0.083. We observe a pronounced decrease of *MR* after annealing at the optimal conditions ($T_a$=280°C), and a substantial increase of *MR* after annealing at a higher temperature. For samples with lower Mn content ($x$~0.03) the *MR* is not affected by the heat treatment at the optimal conditions. Our previous papers [3,5] showed that for low Mn concentration the influence of annealing procedure on both the Curie temperature and conductivity is weak.

In order to study the magnetic properties of the GaMnAs layers we performed magnetooptical Kerr effect (MOKE) measurements. The experiments were carried out in the polar configuration under pulsed magnetic fields up to 25T and at temperatures ranging from 5K up to 250K for epilayers in a wide range of Mn content between x=0.014 and x=0.083. Fig.2 presents Kerr rotation angle $\theta_K$ versus magnetic field up to 8T for the as grown sample with $x$=0.083 and a Curie temperature $T_C$=88K. The data were collected at various temperatures. Additionally, the absolute magnetization *M* of this sample was measured with the use of a SQUID magnetometer (at 5K and magnetic fields up to 5T in the perpendicular configuration). Both MOKE and SQUID measurements indicate that even at low temperatures and high magnetic fields the measured magnetization is far from being saturated. In addition, comparison of the two data sets (Kerr rotation angle $\theta_K$ with the SQUID magnetization *M*) collected in the same field range indicates that, in contrast to metallic magnetic films, the usual relation $\theta_K \propto M$ is not valid for GaMnAs. In the range of low magnetic fields, below 1T and for $T$ below $T_C$, $\theta_K(H)$ exhibits a non monotonic field dependence. This nonmonotonic behavior is observed in the Kerr rotation curves for all of the investigated samples. The origin of the observed feature is not fully understood; however, it is probably related to the domain structure of the investigated material, as suggested by our studies of annealing-induced changes in the strain conditions. Strain relations in the examined samples were measured using high resolution X-ray diffraction method (Philips X'Pert – MRD diffractometer). The layers were found to be fully strained, as indicated by reciprocal lattice maps of the asymmetric 224 reflection, collected for each sample. After annealing, the angular position of the 004 reflection peak for the layer changed (Fig. 3). In Tab.1 we present the calculated values of the relaxed mismatch $(a_{Lrelax}-a_S)/a_S$ before and after annaeling.

---

[*] Corresponding author. Tel.: (+48-22) 843-56-26; fax: (+48-22) 843-09-26; e-mail: kuryl@ifpan.edu.pl.

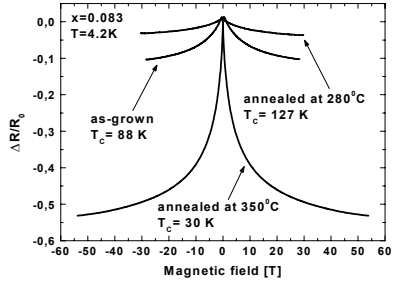

Fig. 1. Magnetoresistivity $(R-R_0)/R_0$, where $R_0$ is the value of resistivity at $B=0$, for the epilayer with high Mn content x=0.083, as-grown, annealed at the optimal conditions ($T_a=280^0$C) and after annealing at higher temperature ($T_a=350^0$C).

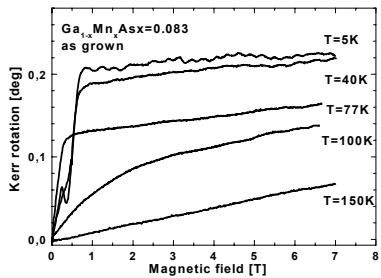

Fig. 2. Kerr rotation angle versus magnetic foield $\theta_K(B)$ for the as-grown $Ga_{0.917}Mn_{0.083}As$ epilayer measured at various temperatures.

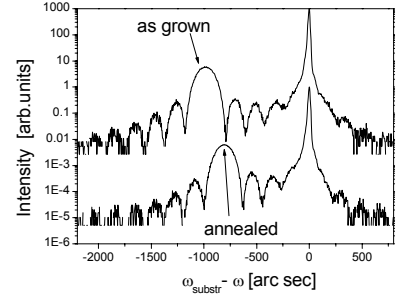

Fig.3. Diffraction curve $\varpi/2\theta$ scan for the 004 reflection.

Tab.1 Relaxed mismatch $(a_{Lrelax}-a_S)/a_S$ before and after annealing at the optimal conditions ($280^0$C), (*ppm-part per million*)

| x | Δa/a – as grown | Δa/a - after anneal. |
|---|---|---|
| 0.083 | 3833 | 3225 |
| 0.061 | 2875 | 2531 |
| 0.027 | 1494 | 1370 |

At present, only a very speculative interpretation of the reported results is possible. We have shown that LT annealing affects the strain of the measured epilayers. This can modify the domain structure, that may in turn be responsible for the $\theta_K(H)$ dependence observed at low magnetic fields, and may also lead to the observed significant changes in magnetoresistivity.

This work was partially supported within European Community program ICA1-CT-2000-70018 (Center of Excellence CELDIS) and Committee for Scientific Research (Poland) project PBZ/KBN/044/P03/2001.We would like to thank R. Szymczak and M. Baran for performing the SQUID measurements.